\documentclass[12pt,a4paper]{article}

\usepackage{amsfonts,amsmath,amsthm}

\newcommand{\mh}{{\mathcal H}}
\newcommand{\znu}{{\mathbb Z^{\stackrel{\nu}{}}}}
\newcommand{\hll}{H_\Lambda}
\newcommand{\dom}{\,\mathrm{Dom}\,}
\newcommand{\supp}{\,\mathrm{supp}\,}

\newcommand{\fl}{\Phi_\Lambda}
\newcommand{\hlo}{H_{\Lambda,0}}
\newcommand{\supf}{\sup_x\|\phi_x\|}
\newcommand{\thll}{\widetilde{H}_\Lambda}
\newcommand{\ol}{\Omega_\Lambda}
\newcommand{\olo}{\Omega_{\Lambda,0}}
\newcommand{\okl}{{\emptyset\ne K\subset\Lambda}}
\newcommand{\oil}{{\emptyset\ne I\subset\Lambda}}
\newcommand{\tol}{\widetilde{\Omega}_\Lambda}
\newcommand{\spec}{\,\mathrm{Spec}\,}
\newcommand{\mhl}{{\mathcal H}_\Lambda}
\newcommand{\ac}{\mathcal A}
\newcommand{\bhl}{\mathcal B(\mathcal H_\Lambda)}
\newcommand{\cc}{\mathbb C}
\newcommand{\rr}{\mathbb R}
\newcommand{\ttt}{\mathbb T}
\newcommand{\Ran}{\,\mathrm{Ran}\,}
\newcommand{\ku}{\widehat{u}}
\newcommand{\kw}{\widehat{w}}
\newcommand{\kv}{\widehat{v}}

\newcommand{\tfl}{\widetilde{\Phi}_\Lambda}
\newcommand{\thlo}{\widetilde{H}_{\Lambda,0}}
\newcommand{\thl}{\widetilde{H}_{\Lambda}}
\newcommand{\nn}{{|\!|\!|}}
\newcommand{\dist}{\,\mathrm{dist}\,}

\newtheorem{theor}{Theorem}
\newtheorem{lemma}{Lemma}
\begin{document}

\title{Quasi-particles in weak perturbations of non-interacting quantum lattice systems}

\author{D.A.Yarotsky \footnote{Department of Mathematical Physics,
University College Dublin, Ireland; e-mail: yarotsky@mail.ru}
\footnote{on leave from Institute for Information Transmission
Problems, Moscow, Russia }}

\date{}
\maketitle

{\bf Abstract.} We consider a general weak perturbation of a
non-interac\-ting quantum lattice system with a non-degenerate
gapped ground state. We prove that the presence of isolated
eigenvalues in the spectrum of the decoupled model leads to the
existence of quasi-particles in the perturbed model. Also, a
scattering theory for asymptotically free many-particle states is
developed.

\section{Introduction and results}

We consider a quantum system on a lattice, which is a weak
perturbation of a non-interacting system with a non-degenerate
gapped ground state. A rigorous perturbation theory for ground
states in such models was developed in
\cite{A,DK,KT1,KT2,KT,M1,M2,Y1,Y2}. In particular, it is known
that the weakly interacting model has a unique gapped ground
state. In the present paper we establish a quasi-particle picture
in the corresponding sector of the model. We prove that if there
are isolated eigenvalues in the spectrum of the decoupled system,
then the weakly interacting system has particle-like states which
are obtained perturbatively from the eigenvectors of the decoupled
system. We then show that the existence of a quasi-particle
subspace leads to subspaces, describing scattering states of
asymptotically free finite collections of quasi-particles.

An explicit quasi-particle picture for quantum Ising model in a
strong magnetic field was developed by Malyshev in \cite{Ma2} (see
also \cite{Ma1,MM2}). The existence of one-particle subspaces in
more complicated quantum spin systems was later shown in
\cite{AMZ2,ZhKM}. All these results for quantum spin systems rely
on a special ground state renormalization, which makes the
Hamiltonian into a generator of a Markov stochastic process (see
\cite{AMZ1,KM,M,MS,MT,I,Y3} for results on quasi-particles for
Markov processes). It appears, however, that the interpretation of
the renormalized system as a stochastic process limits the
applicability of the method and is not actually necessary for the
study of quasi-particles. Indeed, in the present paper we show
that the quasi-particle picture can be established for a quite
general class of quantum models within the $C^*$-algebraic
framework.

We give now precise definitions.

We consider a quantum ``spin'' system on the lattice $\znu$.
Suppose that for each $x\in\znu$ there is a Hilbert space $\mh_x$
(possibly infinite-dimensional) associated with this site. For the
restriction to  a finite volume $\Lambda\subset \znu$ we will use
the notation $$\mh_\Lambda:=\otimes_{x\in\Lambda}\mh_x.$$
 The (formal) Hamiltonian of the model consists of a non-interacting Hamiltonian
and a perturbation: $$H=H_0+\Phi.$$ Here $H_0$ is the free
Hamiltonian: $$H_0=\sum_{x\in\znu}h_x.$$ We assume that each $h_x$
is a non-negative self-adjoint, possibly unbounded operator on
$\mh_x$ with a non-degenerate ground state $\Omega_x\in\mh_x$:
$$h_x\Omega_x=0$$ and a  spectral gap $\ge 1$:
\begin{equation}\label{gap}
h_x|_{\mh_x\ominus\Omega_x}\ge{\bf 1}\end{equation} (this is
necessary and sufficient in order that the non-interacting
Hamiltonian have a non-degenerate ground state and a spectral gap
$\ge 1$; here and in the sequel we slightly abuse the notation by
denoting the one-dimensional subspace spanned by $\Omega_x$ with
the same symbol). In order to define the perturbation $\Phi$ we
fix a finite subset $\Lambda_0\subset\znu$ (range of the
perturbation) and set
\begin{equation}\label{f}\Phi=\sum_{x\in\znu}\phi_x,\end{equation}
where $\phi_x$ is a self-adjoint bounded operator on
$\mh_{\Lambda_0+x}$ (here $\Lambda_0+x$ is a shift of
$\Lambda_0$). We will assume that the perturbation is small in the
sense that $\sup_{x\in\znu}\|\phi_x\|$ is finite and small enough.

Under these assumptions the model has a unique infinite volume
ground state with a spectral gap and an exponential decay of
correlations . It can be obtained rigorously as the limit of
finite volume ground states. Let $\Lambda\subset\znu$ be a finite
volume and $\hll$ the restriction of the Hamiltonian $H$ to
$\Lambda$ with empty boundary conditions:
$$\hll:=\hlo+\Phi_\Lambda,$$ where
\begin{equation}\label{hlo}
\hlo:=\sum_{x\in\Lambda}h_x,\quad
\Phi_\Lambda:=\sum_{x\in\znu:\Lambda_0+x\subset \Lambda}\phi_x
\end{equation}
(other boundary conditions can be used as well). Since $\fl$ is
bounded, $\hll$ is self-adjoint with $\dom(\hll)=\dom(\hlo)$. The
following theorem shows that the presence of a non-degenerate
ground state and a perturbative spectral estimate hold uniformly
for all finite volumes:

\begin{theor}
There exists a constant $c_1$, depending only on the perturbation
range $\Lambda_0$, such that if $\supf<c_1$ then for any finite
$\Lambda$ the Hamiltonian $\hll$ has a non-degenerate ground state
$\ol$: $$\hll\ol=E_\Lambda\ol,\quad \hll|_{\mhl\ominus\ol}>
E_\Lambda{\bf 1},$$ where $E_\Lambda$ is the ground state energy.
Moreover, let $$\thll:=\hll-E_\Lambda{\bf 1}$$ be the renormalized
Hamiltonian. There exists a constant $c_2=c_2(\Lambda_0)$ such
that
\begin{equation}\label{sl}
\spec(\thll)\subset\bigcup_{a\in\spec(\hlo)}\{z:|z-a|\le c_2\supf
a\}.
\end{equation}
In particular, this gives a lower bound for the gap between 0 and
the rest of the spectrum:
\begin{equation}\label{slg}
\thll|_{\mhl\ominus\ol}\ge(1-c_2\supf){\bf 1}.\end{equation}
\end{theor}

Now the thermodynamic limit of the ground states can be performed;
the limiting, infinite volume ground state is to be understood as
a state ($\equiv$ normalized positive linear functional) on the
algebra of local observables (see, e.g, \cite{BR}). Let $\bhl$ be
the algebra of bounded operators in $\mhl$ for any finite
$\Lambda$, and
$$\ac_\infty:=\bigcup_{\Lambda\subset\znu,|\Lambda|<\infty}\bhl$$
the full local algebra. Let $\Lambda\nearrow\znu$ mean that
$\Lambda$ converges to $\znu$ in the sense that it eventually
contains any finite subset. Then

\begin{theor}
There exists the thermodynamic week$^*$-limit $\omega_\infty$ of
the finite volume ground states: $$\langle A\ol,\ol\rangle
\xrightarrow{\Lambda\nearrow\znu}\omega_\infty(A), \quad
A\in\ac_\infty.$$
\end{theor}

Throughout the paper by $\langle\cdot,\cdot\rangle$ we denote
various scalar products.

Using appropriate definition of the infinite volume ground state,
one can show that in our model it is unique: see \cite{Y2}.

Having found the infinite volume ground state, we use the GNS
construction to define in a conventional way the Hilbert space
$\mh_\infty$, describing local excitations of this ground state.
Precisely, we consider  the GNS triple
$(\mh_\infty,\pi_\infty,\Omega_\infty)$ (Hilbert space, cyclic
representation of $\ac_\infty$ in $\mh_\infty$, and the cyclic
vector), associated with the state $\omega_\infty$ so that
$$\langle\pi_\infty(A)\Omega_\infty,\Omega_\infty\rangle
=\omega_\infty(A).$$ Next we define the infinite volume
Hamiltonian $H_\infty$ on $\mh_\infty$ as an appropriate limit of
$\thll$; it is convenient to use the (week) resolvent convergence,
because resolvent expansions are very relevant for the
perturbation theory.

\begin{theor}
There exists a (unique) self-adjoint operator $H_\infty$ on
$\mh_\infty$, which is the weak resolvent limit of $\thll$ in the
following sense: for any $A,B\in\ac_\infty$ and
$z\in\cc\setminus\rr$
\begin{equation}\label{lt}
\langle(\thll-z)^{-1}A\Omega_{\Lambda},B\Omega_{\Lambda}\rangle
\xrightarrow{\Lambda\nearrow\znu}
\langle(H_\infty-z)^{-1}\pi_\infty(A)\Omega_\infty,\pi_\infty(B)\Omega_\infty\rangle
.\end{equation} Moreover, $H_\infty\Omega_\infty=0$ and estimates
(\ref{sl}),(\ref{slg}) hold with $\thll,\mhl,\ol$ replaced by
$H_\infty,\mh_\infty,\Omega_\infty$, and $\spec(\hlo)$ replaced by
$$\spec(H_{\infty,0}):=\Bigl\{\sum_{x\in\znu}a_x\Bigl|a_x\in\spec(h_x),a_x\ne
0 \text{ only for finitely many } x\Bigr\}.$$
\end{theor}

Theorems 1-3 have been proved in \cite{A,KT2,Y1}.

We are now in a position to discuss the main topic of this paper,
the quasi-particle excitations. In addition to the assumptions
made above we assume now that the model is translationally
invariant, i.e., $\mh_x,h_x,\phi_x,\Omega_x$ are translates of
some $\mh,h,\phi,\Omega$. This implies that the ground state is
translationally invariant and there is a representation of lattice
shifts by unitary operators $U_x:\mh_\infty\to
\mh_\infty,x\in\znu$, commuting with the Hamiltonian $H_\infty$.
Let $\ttt^{\nu}$ be the $\nu$-dimensional torus of quasi-momenta
(we will identify $\ttt$ with the segment $[0,1]$) and consider
the Hilbert space $L_2(\ttt^{\nu})$. We shall say that an
invariant with respect to $H$ and $U_x$ subspace $\mathcal
G\subset\mh_\infty$ is  {\bf a one-particle subspace} if there is
a unitary $V:\mathcal G\to L_2(\ttt^{\nu})$ such that $VU_xV^*$ is
the multiplication by $ e^{2\pi i\langle x,\cdot\rangle}$ and
$VH_\infty V^*$ the multiplication by some function $m(\cdot)$
(which is the energy-momentum relation of the quasi-particle).
Equivalently, an invariant $\mathcal G$ is a one-particle subspace
if there exists an orthonormal basis $\xi_x,x\in\znu,$ in
$\mathcal G$ such that $U_x\xi_y=\xi_{x+y}$; the unitary $V$ is
then given by $$V:\xi_x\mapsto e^{2\pi i\langle x,\cdot\rangle}.$$
The natural situation in which one expects the presence of a
one-particle subspace and which we will only deal with is when the
non-interacting Hamiltonian $H_{\infty,0}$ has a spectrally
isolated one-particle subspace and the perturbation is
sufficiently small. We assume therefore now that the single-site
Hamiltonian $h$ has an isolated non-degenerate eigenvalue $\mu$
and
$\mu\notin\cup_{k>1}\{\mu_1+\ldots+\mu_k|\mu_i\in\spec(h),\mu_i\ne
0\},$ so that $\mu$ is an isolated eigenvalue of $H_{\infty,0}$,
corresponding to the one-particle subspace spanned by
$w_x\otimes(\otimes_{y\ne x}\Omega_y),x\in\znu$, where $w$ is the
eigenvector of $h$: $$hw=\mu w;$$ the function $m$ identically
equals $\mu$ in this case. By Th.3 and (\ref{sl}), if $\|\phi\|$
is sufficiently small then the part of the spectrum of $H_\infty$,
lying in the segment $[(1-c_2\|\phi\|)\mu,(1-c_2\|\phi\|)\mu]$, is
separated by gaps from the rest of the spectrum. Let $\mh_1$ be
the spectral subspace of $H_\infty$, corresponding to this
segment. Then

\begin{theor}
If $\|\phi\|$ is sufficiently small, then $\mh_1$ is a
one-particle subspace, and the corresponding energy-momentum
relation  $m$ is a real analytic function.
\end{theor}

The final result we discuss here concerns existence of
asymptotically free  many-particle scattering states. Let
$\mathcal F=\mathcal F(\mh_1)$ be the symmetric Fock space
obtained from the above one-particle subspace $\mh_1$ and
describing various finite collections of these quasi-particles (in
fact, though the indistinguishability of the particles is
essential, the particular form of statistics is not relevant for
the result below). The free evolution of a finite collection of
quasi-particles is governed by the Hamiltonian $H_f$, acting on
$\mathcal F$ and defined as the second quantization of
$H_\infty|_{\mh_1}$. We will show that the initial Hilbert space
$\mh_\infty$ contains invariant subspaces, describing
asymptotically, in distant past or future, free collections of
quasi-particles. We have to assume now that the energy momentum
relation $m$ is non-constant, otherwise there is no scattering and
the particle picture breaks down. Denote by $U_{f,x}:\mathcal
F\to\mathcal F$ the second quantization of the unitary lattice
shifts $U_x|_{\mh_1}$ in the one-particle subspace. Then

\begin{theor}
There exist isometric wave operators $W_{\pm}:\mathcal F\to
\mh_\infty$ such that $\Ran(W_{\pm})$ are invariant subspaces of
$H_\infty,U_x$, and $H_\infty|_{{\rm Ran}(W_{\pm})} =W_{\pm} H_f
W_{\pm}^*$, $U_x|_{{\rm Ran}(W_{\pm})} =W_{\pm}U_{f,x}W_{\pm}^*$.
\end{theor}

Our proofs of Theorems 4 and 5 closely follow the paper \cite{Y3}
and we will omit here some technical details. The proof of Theorem
4 is a modification of Malyshev-Minlos technique (see
\cite{AMZ1,AMZ2,I,KM,MM2,M,MS,MT,Y3,ZhKM}). The proof of Theorem 5
follows general ideas of scattering theory for quantum many-body
systems. In \cite{Ma2}, the scattering theory was based on the
well-known construction of  many-particle states given by Haag and
Ruelle in the axiomatic quantum field theory \cite{H1,H2,R}. Our
exposition is closer to the  spin wave scattering for the
Heisenberg ferromagnet as presented in \cite{RS3}, section XI.14.

\section{Preliminary results. Theorems 1-3}

All theorems stated above rely heavily on perturbative expansions
for the ground state, one-particle states and the Hamiltonian. We
begin by briefly reviewing the proofs of Ths.1-3 (see \cite{Y1}
for details). From now on we adopt for brevity the following
convention. We will denote by $c$ and $\epsilon$ various
(generally different in different formulas) positive constants,
which do not depend on the volume $\Lambda$, though may depend on
the interaction range $\Lambda_0$. We write $\epsilon$ if this
constant can be chosen arbitrarily small by choosing $\supf$ small
enough; on the other hand the constant $c$ is typically greater
than 1 and does not depend on $\supf$.

The ground state vector $\Omega_\Lambda$ of the Hamiltonian
$H_\Lambda$ in a finite volume can be found using a suitable
ansatz, reflecting smallness of correlations between distant
spins.

For any $I\subset\Lambda$ set $$\mh_I':=\otimes_{x\in
I}\mh_x',\;\;\Omega_{I,0}:=\otimes_{x\in  I}\Omega_x$$ (with
$\mh'_\emptyset\equiv\cc$). It follows that
\begin{equation}\label{mmo}\mhl\ominus\olo=\bigoplus_\oil\mh_I'\otimes
\Omega_{\Lambda\setminus I,0}.\end{equation} We will typically
denote vectors from $\mh_I'$ by $u_I,v_I$, etc. For each
$u_I\in\mh_I'$ we introduce a ``creation'' (or ``spin raising'')
operator $\ku_I$ in $\mh_I$ by $$\ku_Iv=\langle
v,\Omega_{I,0}\rangle u_I,\;\;\;v\in\mh_I$$ (with $\ku_\emptyset$
a scalar operator). A useful property of these operators is that
for any $I$, $J$ and $u_I,v_J$ $$\ku_I\kv_J=\left\{
\begin{array}{ll}0&\text{ if }I\cap J\ne\emptyset, \\
\widehat{u_I\otimes v_J}&\text{ if }I\cap J=\emptyset.
\end{array}\right.$$ In particular, they commute. For any
$v\in\mhl$ such that $\langle v,\olo\rangle=1$ there exists a
unique collection $\{v_I\in\mh_I'\}_\oil$ such that
$\exp\bigl(\sum_\oil\kv_I\bigr)\olo=v$ (these $v_I$ can be
obtained by truncation from components of $v$ appearing in the
decomposition (\ref{mmo})). Let $\tol$ be the ground state vector
of $H_\Lambda$ normalized so that $\langle \tol,\olo\rangle=1$
(i.e., $\tol:=\Omega_\Lambda/\langle\Omega_\Lambda,\olo\rangle$).
Initially the non-degeneracy  of the ground state and its
non-orthogonality to $\olo$ is clear from the usual finite-volume
perturbation theory for sufficiently weak perturbations in each
particular volume, and it can be shown that the estimate for the
perturbation which ensures this property can actually be chosen
uniform in the volume. Let $\{v^{(\Lambda,gs)}_I\in\mh_I'\}_\oil$
be the corresponding collection such that
$$\tol=\exp\bigl(\sum_\oil\kv^{(\Lambda,gs)}_I\bigr)\olo.$$

\begin{lemma}
For any $\epsilon>0$, if $\supf$ is sufficiently small then
\begin{equation}\label{c1}
\max_{x\in\Lambda}\sum_{I\subset\Lambda: x\in
I}\|H_{I,0}v_I^{(\Lambda,gs)}\|\epsilon^{-(d_I+1)}\le
1,\end{equation} where $d_I$ is the minimal length of a connected
graph containing $I$.
\end{lemma}
The idea of the proof is to rewrite the Schr\"odinger equation
$H_\Lambda\tol=E_\Lambda\tol$ as a fixed point equation for a
suitable (non-linear) mapping on the space of collections, which
is then shown to be a contraction in the set specified by
Eq.(\ref{c1}).

Having found the ground state $\tol$, it is convenient for the
study of the operator $H_\Lambda$ to use for vectors $v\in\mhl$
the expansion
\begin{equation}\label{vsum}v=\sum_{I\subset\Lambda}\kv_I\tol.
\end{equation}
For any $v$, there exists a unique collection such that
(\ref{vsum}) holds: indeed, if we introduce the operators
$P_{\Lambda,I}:\mhl\to\mh_I'$ by
$$u=\sum_{I\subset\Lambda}(P_{\Lambda,
I}u)\otimes\Omega_{\Lambda\setminus I,0}, \quad \forall
u\in\mhl,$$ then
\begin{eqnarray}\label{p}
v_K & = &
P_{\Lambda,K}\sum_{J\subset\Lambda}\kv_J\Omega_{\Lambda,0}
\nonumber
\\ & = & P_{\Lambda,K}\exp\bigl(-\sum_\oil\kv^{(\Lambda,gs)}_I\bigr)
\sum_{J\subset\Lambda}\kv_J
\exp\bigl(\sum_\oil\kv^{(\Lambda,gs)}_I\bigr) \olo \nonumber  \\ &
= & P_{\Lambda,K}\exp\bigl(-\sum_\oil\kv^{(\Lambda,gs)}_I\bigr) v.
\end{eqnarray}
Using expansion (\ref{vsum}), we rewrite the renormalized
Hamiltonian $\thl$ as
\begin{equation}\label{thl}\thl=\thlo+\tfl,\end{equation} where $\thlo$ is the
``diagonal'' part defined by
\begin{equation}\label{thlo}\thlo\kv_I\tol=\widehat{H_{I,0}v_I}\tol.
\end{equation} For the
renormalized perturbation $\tfl$ we then have
\begin{eqnarray*}
\tfl\kv_I\tol & = & (\thl-\thlo)\kv_I\tol \\ & = &
\thl\kv_I\tol-\widehat{H_{I,0}v_I}\tol \\ & = &
[\thl,\kv_I]\tol-\widehat{H_{I,0}v_I}\tol \\ & = &
[\hlo,\kv_I]\tol+[\fl,\kv_I]\tol-\widehat{H_{I,0}v_I}\tol \\ & = &
[\fl,\kv_I]\tol.
\end{eqnarray*}
It is convenient to write the operator $\tfl$ in the form
\begin{equation}\label{tflk}
\tfl\kv_I\tol=\sum_{J\subset\Lambda}\widehat{(F_\Lambda
v_I)_J}\tol\end{equation} with some $(F_\Lambda v_I)_J\in\mh'_J$,
thereby expanding the image vector as in  (\ref{vsum}). Using
(\ref{p}), we find that
\begin{eqnarray*}
(F_\Lambda v_I)_J & = & P_{\Lambda,
J}\exp\bigl(-\sum_\okl\kv^{(\Lambda,gs)}_K\bigr)[\Phi_\Lambda,\kv_I]\tol
\\ & = & P_{\Lambda,
J}\exp\bigl(-\sum_\okl\kv^{(\Lambda,gs)}_K\bigr)[\Phi_\Lambda,\kv_I]
\exp\bigl(\sum_\okl\kv^{(\Lambda,gs)}_K\bigr)\Omega_{\Lambda,0}.
\end{eqnarray*}
Expanding this expression into a commutator series and using Lemma
1, one finds that \begin{equation}\label{sumj}\sum_{J\subset
\Lambda}\|(F_\Lambda v_I)_J\|\epsilon^{-(d_{J;I}+1)}\le c \supf
|I| \|v_I\|,\end{equation} where by $d_{J;I}$ we denote the
minimal length of a graph connecting all sites in $J$ to some
sites in $I$. In particular, this estimate shows that $\tfl$ is a
relatively bounded, in some special sense, perturbation of
$\thlo$. Indeed, we introduce a new norm $\nn\cdot\nn$ in
$\mh_\Lambda$ by
$$\nn\sum_{I\subset\Lambda}\ku_I\tol\nn:=\sum_{I\subset\Lambda}\|u_I\|.$$
Then by (\ref{sumj}) for any $v\in\mh_\Lambda$ $$\nn\tfl v\nn\le
c\supf\nn\thlo v\nn.$$ We assume that $\supf$ is small so that
$c\supf<1$. If $z\in\cc$ lies outside of the union of circles
standing on the r.h.s. of (\ref{sl}), then
$\nn\tfl(\thlo-z)^{-1}\nn<1$ and hence the resolvent
$(\thl-z)^{-1}$ exists and is given by the exponentially
convergent series
\begin{equation}\label{t-z}
(\thlo-z)^{-1}\sum_{k=0}^\infty(-1)^k(\tfl(\thlo-z)^{-1})^k.
\end{equation} That proves the spectral estimate (\ref{sl}).

We consider now the thermodynamic limit $\Lambda\nearrow\znu$. It
easily follows from the proof of Lemma 1 that the quantities
$v_I^{(\Lambda,gs)}$, defining the ground state in finite volumes,
have limits
$$v_I^{(\infty,gs)}:=\lim_{\Lambda\nearrow\znu}v_I^{(\Lambda,gs)}.$$
Using this, Theorem 2 follows immediately by cluster expansions.
As a by-product of cluster expansions, one obtains an exponential
decay of correlations in the ground state:
\begin{equation}\label{dc}|\omega_\infty(A_1A_2)-\omega_\infty(A_1)\omega_\infty(A_2)|\le
c^{|\Lambda_1|+|\Lambda_2|}\epsilon^{\dist(\Lambda_1,\Lambda_2)}\|A_1\|\|A_2\|,\;\;\;A_i\in\mathcal
B(\mh_{\Lambda_i})\end{equation} Iterating this inequality, one
finds that
\begin{eqnarray}\label{oia}&  &|\omega_\infty(A_1\cdots
A_n)-\omega_\infty(A_1)\cdots\omega_\infty(A_n)|  \\ & \le &
(n-1)c^{|\Lambda_1|+\ldots+|\Lambda_n|}\epsilon^{\min_{i\ne
j}\dist(\Lambda_i,\Lambda_j)}, \qquad  A_i\in\mathcal
B(\mh_{\Lambda_i}).\nonumber
\end{eqnarray}

Now we discuss the thermodynamic limit of the Hamiltonian. Recall
that $(\mh_\infty,\pi_\infty,\Omega_\infty)$ is the GNS triple
associated with the ground state $\omega_\infty$. The limiting
Hamiltonian $H_\infty$ is conventionally defined on the vectors of
the form $\pi_\infty(A)\Omega_\infty$, with $A\in\mathcal
A_\infty$ such that $[H,A]\in\mathcal A_\infty$, by
$$H_\infty\pi_\infty(A)\Omega_\infty:=\pi_\infty([H,A])\Omega_\infty.$$
It is convenient to take $A=\ku_I$ here. Consider the space
$\mathcal U$ of finite linear combinations of operators $\ku_I$.
It is easy to see that the set
$\pi_\infty(\mathcal{U})\Omega_\infty$ is dense in $\mh_\infty$.
Extending the formulas (\ref{thl}),(\ref{thlo}),(\ref{tflk}), we
have
\begin{equation}\label{hi}
H_\infty\pi_\infty(\ku_I)\Omega_\infty=
\pi_\infty(\widehat{H_{I,0}u _I})\Omega_\infty+
\pi_\infty\Bigl(\sum_{J\subset\znu,|J|<\infty}\widehat{(F_\infty
u_I)_J}\Bigr)\Omega_\infty,
\end{equation}
where $$(F_\infty u_I)_J:=\lim_{\Lambda\nearrow\znu}(F_\Lambda
u_I)_J.$$ By this formula the operator $H_\infty$ is densely
defined on the subspace spanned by vectors
$\pi_\infty(\ku_I)\Omega_\infty$ with
$u_I\in\mh_I'\cap\dom(H_{I,0})$. The estimate (\ref{sumj}) remains
valid for the limiting quantities $(F_\infty u_I)_J$, so the
resolvent expansion of the form (\ref{t-z}) for
$(H_\infty-z)^{-1}$, with the same values of $z$ as before, is
exponentially convergent on vectors from $\pi_\infty(\mathcal
U)\Omega_\infty$. The essential self-adjointness of $H_\infty$ and
the resolvent convergence (\ref{lt}) follow by standard arguments.

\section{Proof of Theorem 4}

The orthogonal projector onto the spectral subspace $\mh_1$,
corresponding to the neighborhood of the point $\mu$ as defined in
the Introduction, is given by $$P_{\mh_1}=-(2\pi
i)^{-1}\int_\gamma(H_\infty-z)^{-1}dz,$$ where $\gamma$ is a
contour in the complex plane  surrounding this neighborhood. Since
$\gamma$ lies outside of the union of circles standing on the
r.h.s. of (\ref{sl}), we can use the resolvent expansion
(\ref{t-z}) to estimate projections of vectors
$\ku_I\Omega_\infty$. For the normalized eigenvector $w_x$ of $h$
at the site $x$, using estimate (\ref{sumj}) we obtain
\begin{equation}\label{pmw}
P_{\mh_1}\kw_x\Omega_\infty=\kw_x\Omega_\infty+\sum_I\kv_I^{(w_x)}\Omega_\infty,
\end{equation} with some vectors $v_I^{(w_x)}$ such that
\begin{equation}\label{pmwest}
\sum_I\|H_{I,0}v_I^{(w_x)}\|\epsilon^{-(d_{I\cup\{x\}}+1)}\le
1.\end{equation} The first term in (\ref{pmw}) comes from the
leading term in the resolvent expansion; the second term, coming
from the rest of the expansion, is small due to (\ref{sumj}). It
is easy to see that the vectors $P_{\mh_1}\kw_x\Omega_\infty$ are
total in $\mh_1$. Indeed, since $\mathcal U\Omega_\infty$ is dense
in $\mh_\infty$, it suffices to show that for any $u_I$ we can
expand $P_{\mh_1}\ku_I\Omega_\infty$ as a sum of
$P_{\mh_1}\kw_x\Omega_\infty,x\in\znu$, with some coefficients.
For any $u_I$, if $|I|>1$, or if $I=\{x\}$ and $\langle
u_x,w_x\rangle=0$ (let us call for the moment such vectors
$\ku_I\Omega_\infty$ `irrelevant', as opposed to the `relevant'
$\kw_x\Omega_\infty$), we have an expansion similar to
(\ref{pmw}), but with the vanishing leading term:
\begin{equation}\label{pku}
P_{\mh_1}\ku_I\Omega_\infty=\sum_J\kv_J^{(u_I)}\Omega_\infty,
\end{equation} where
$$\sum_J\|H_{J,0}v_J^{(u_I)}\|\epsilon^{-(d_{J;I}+1)}\le
\|u_I\|.$$ In particular, $\sum_J\|v_J^{(u_I)}\|\le \|u_I\|/2$ for
such $u_I$. Now, on the r.h.s. of (\ref{pku}), we single out the
contribution spanned by $\kw_x\Omega_\infty,x\in\znu,$ and apply
$P_{\mh_1}$ to both sides of the equality. Since
$P_{\mh_1}^2=P_{\mh_1}$, we see that the projection of an
irrelevant vector is represented as a sum of relevant projections
plus projections of irrelevant vectors with a smaller total norm;
the iteration of this procedure yields the desired expansion of
$P_{\mh_1}\ku_I\Omega_\infty$ in terms of
$P_{\mh_1}\kw_x\Omega_\infty$.

Now we orthogonalize the vectors
$P_{\mh_1}\kw_x\Omega_\infty,x\in\znu$. Let $G$ be the Gram matrix
of this family: $$G_{xy}=\langle P_{\mh_1}\kw_x\Omega_\infty,
P_{\mh_1}\kw_y\Omega_\infty\rangle. $$ Using the decay of
correlation estimate (\ref{dc}) and the expansion (\ref{pmw}) with
the estimate (\ref{pmwest}), one finds that
\begin{equation}\label{gram}
|G_{xy}-\delta_{xy}|\le\epsilon^{|x-y|+1}.
\end{equation}
We obtain an orthogonal family $\xi_x$ by applying $G^{-1/2}$ to
the vectors $P_{\mh_1}\kw_x\Omega_\infty$:
$$\xi_x=\sum_yG^{-1/2}_{xy}P_{\mh_1}\kw_y\Omega_\infty.$$ It is
easy to see that the exponential estimate (\ref{gram}) for $G$
implies similar estimate for $G^{-1/2}$; as a result, we have for
$\xi_x$ an expansion analogous to (\ref{pmw}) with an exponential
estimate analogous to (\ref{pmwest}):
\begin{equation}\label{pmw1}
\xi_x=\kw_x\Omega_\infty+\sum_I\kv_I^{(\xi_x)}\Omega_\infty,
\end{equation}
where
\begin{equation}\label{pmwest1}
\sum_I\|H_{I,0}v_I^{(\xi_x)}\|\epsilon^{-(d_{I\cup\{x\}}+1)}\le
1.\end{equation} We have thus shown that $\mh_1$ is a one-particle
subspace, and it only remains to check that the energy-momentum
relation is analytic. It follows from the exponential decay of
$\langle H_\infty\xi_x,\xi_y\rangle$ at $|x-y|\to\infty$, which in
turn follows from the above expansion (\ref{pmw1}),(\ref{pmwest1})
using decay of correlations (\ref{dc}).

\section{Proof of Theorem 5}

An important point in the proof of Theorem 5 is the observation
that distant excitations of the ground state evolve independently.
Recall that the Hamiltonian $H_\infty$ is defined by $H_\infty
A\Omega_\infty=[H,A]\Omega_\infty,$ where $A$ is any local
operator such that $[H,A]$ is bounded. If operators $A_i$ act on $
\mh_{I_i}, i=1,\ldots,n,$ and the distance between different
$I_i$'s is greater than the range of interaction, then
$$[H,A_1A_2\cdots A_n]=A_2\cdots A_n[H,A_1]+\ldots+A_1\cdots
A_{n-1}[H,A_n].$$ In particular, if $A_i=\kv_{I_i}$ with some
$v_{I_i}\in \mh_{I_i}'\cap\dom(H_{I_i,0})$, where $I_i$'s are
separated by distances not less than the interaction range, then
\begin{equation}\label{hkk}
H_\infty\kv_{I_1}\kv_{I_2}\cdots\kv_{I_n}\Omega_\infty=
\kv_{I_2}\cdots\kv_{I_n}H_\infty\kv_{I_1}\Omega_\infty+\ldots+
\kv_{I_1}\cdots\kv_{I_{n-1}}H_\infty\kv_{I_n}\Omega_\infty.
\end{equation}
In the previous section we found a basis $\xi_x$ in the
one-particle subspace $\mh_1$, whose vectors have the
well-localized expansion (\ref{pmw1}),(\ref{pmwest1}). The time
evolution of a quasi-particle state is straightforwardly described
in the momentum picture, where such a state is identified with a
function on the torus $\ttt^\nu$ of quasi-momenta. In this picture
we can form one-particle states, propagating as wave packets in
certain directions, and, using (\ref{pmwest1}), these states can
be shown to be spatially well-localized. If we consider a few such
wave packets, moving in different direction away from each other,
then in the distant past or future they will form far separated
excitations of the ground state. By (\ref{hkk}), the evolution of
the corresponding ``product'' state in $\mh_\infty$ is then
asymptotically close to the product of respective one-particle
evolutions. Precisely this relation can be stated as the existence
of isometric wave operators intertwining the second quantization
of the one particle evolution with the evolution generated by
$H_\infty$ in a subspace of $\mh_\infty$.

Following the standard setting of scattering theory for a pair of
Hilbert spaces (see \cite{RS3}), we begin by introducing the
operator $T:\mathcal F\to \mathcal H_\infty$, defining an
approximate ``product'' of one-particle states. Let $\mathcal F_n$
be the $n$-particle subspace of the bosonic Fock space $\mathcal
F$. The normalized symmetrized products
$(\xi_{x_1}\otimes\cdots\otimes\xi_{x_n})^{sym}$ form an
orthonormal basis in $\mathcal F_n$. Using expansion (\ref{pmw1}),
we set $$T:(\xi_{x_1}\otimes\cdots\otimes\xi_{x_n})^{sym}\mapsto
\prod_{k=1}^n
(\kw_{x_k}+\sum_I\kv_I^{(\xi_{x_k})})\Omega_\infty.$$ This
operator does not generally extend to a bounded operator on
$\mathcal F$, but it is well-defined on vectors of the form
$\sum_{n=1}^Nk_{x_1,\ldots,x_n}
(\xi_{x_1}\otimes\cdots\otimes\xi_{x_n})^{sym}$, where $N$ is
finite and $\sum_{n=1}^N|k_{x_1,\ldots,x_n}|<\infty$; this will be
sufficient for us. In the Fock space $\mathcal F$ we find a dense
subset $S$ such that the quantities $$W_t{\bf
f}:=\exp\{itH_\infty\}T\exp\{-itH_f\}{\bf f}$$ are well-defined
for ${\bf f}\in S$ and $t\in \rr$, and there exist isometric
limits \begin{equation}\label{wo}W_\pm{\bf
f}:=\lim_{t\to\pm\infty}W_t{\bf f}.\end{equation} Then
$W_\pm:\mathcal F\to H_\infty$ are the desired isometric wave
operators, implementing the intertwining relation $W_\pm H_f=
H_\infty W_\pm$. The intertwining relation for lattice
translations, $W_\pm U_{f,x}= U_x W_\pm$, is obvious because
$TU_{f,x}= U_xT$.

The dense set $S$ is spanned by collections of quasi-particles
forming smooth wave packets drifting away from each other.
Precisely, recall that by definition of the one-particle subspace
we have a unitary $V:\mh_1\to L_2(\ttt^\nu)$ defining the momentum
picture. If $f\in\mh_1$ is such that $Vf\in C^\infty(\ttt^\nu)$,
then a standard stationary phase estimate shows that, roughly
speaking, this wave packet propagates with velocities from the set
$\{2\pi\nabla m(p)|p\in\supp (Vf)\}$:
\begin{lemma}[\cite{RS3}]
Let $Vf \in C^\infty(\ttt^\nu)$ and $\mathcal O\subset\rr^\nu$ be
an open set containing $\{2\pi\nabla m(p)|p\in\supp (Vf)\}$. Then
for arbitrarily large $a$ there exists a constant $c=c(f,\mathcal
O,a)$ such that $$|\langle
\exp\{itH_\infty\}f,\xi_x\rangle|=\Bigl|\int_{\ttt^{\nu}}e^{itm(p)-2\pi
i\langle x,p\rangle}Vf(p)dp\Bigr|\le c(1+|x|+|t|)^{-a}$$ for
$x/t\notin \mathcal O$.
\end{lemma}
This Lemma shows that if $\exp\{itH_\infty\}f$ is expanded as
$\sum_{x\in\znu} k_x\xi_x$, then the coefficients $k_x$ are small
for $x\notin\{2\pi t\nabla m(p)|p\in\supp(Vf)\}$.

Consider now the subset $S_n\subset\mathcal F_n$ formed by vectors
$(f_1\otimes\cdots\otimes f_n)^{sym}$ such that $Vf_k$ are
$C^\infty$ and $$\nabla m(p_k)\ne \nabla m(p_l)\text{ for
}p_k\in\supp(Vf_k),p_l\in\supp(Vf_l),k\ne l.$$ We define $S$ as
the linear span of $\cup_n S_n$. Since the energy-momentum
relation $m(p)$ is a non-constant analytic function, it is easy to
see that $S_n$ is a total subset of $\mathcal F_n$ and hence $S$
is dense in $\mathcal F$.

We will show now that the wave operators (\ref{wo}) exist for
${\bf f}\in S_n$ (and hence for ${\bf f}\in S$). By the Cook
method, it suffices to prove that
\begin{equation}\label{cook}
\int_\rr\|(H_\infty T-TH_f)\exp\{-itH_f\}{\bf
f}\|dt<\infty.\end{equation} The question is whether the integrand
falls off fast enough as $t\to\infty$. We will see that it is
actually $o(|t|^a)$ for any $a$.

 Note first that for any
$f\in\mh_1$ we can expand $f_t:=\exp\{-itH_\infty\}f$ as
\begin{equation}\label{ft}
f_t=\sum_I\kv_I^{(f_t)}\Omega_\infty \end{equation} by first
expanding $f_t$ in $\xi_x$ and then expanding the latter as in
(\ref{pmw1}). If $Vf\in C^\infty(\ttt^\nu)$ and $\mathcal O$ is an
open vicinity of $\{2\pi\nabla m(p)|p\in\supp (Vf)\}$, then it
follows from Lemma 3 and estimate (\ref{pmwest1}) that in
(\ref{ft}) the contribution from the terms with $I\not\subset
t\mathcal O$ is small: precisely, for any $a$
\begin{equation}\label{re1}\sum_{I\not\subset t\mathcal
O}\|H_{I,0}v_I^{(f_t)}\|=o(|t|^a),\text{ as }t\to\infty.
\end{equation}
Moreover, if $g_t:=H_\infty\sum_{I\not\subset t\mathcal
O}\kv_I^{(f_t)}\Omega_\infty$, then by (\ref{hi}) and estimate
(\ref{sumj}) we have that $g_t=\sum_I\kv_I^{(g_t)}\Omega_\infty$
with
\begin{equation}\label{re2}\sum_I\|v_I^{(g_t)}\|=o(|t|^a).
\end{equation} Now
suppose that ${\bf f}\in S_n$, $ { \bf f}=(f_1\otimes\cdots\otimes
f_n)^{sym}$. Then $\exp\{-itH_f\}{\bf
f}=(f_{1,t}\otimes\cdots\otimes f_{n,t})^{sym}$, where
$f_{k,t}:=\exp\{-itH_\infty\}f_k$. Choose the corresponding open
sets $\mathcal O_k,k=1,\ldots,n,$ so that they don't overlap. Then
using (\ref{re1}) one finds that
\begin{equation}\label{ht}
H_\infty T\exp\{-itH_f\}{\bf f}=H_\infty\sum_{I_1\subset t\mathcal
O_1}\kv_{I_1}^{(f_{1,t})}\cdots\sum_{I_n\subset t\mathcal
O_n}\kv_{I_n}^{(f_{n,t})}\Omega_\infty+o(|t|^a)
\end{equation}
for any $a$. On the other hand, using (\ref{re2}) one finds that
\begin{eqnarray}\label{th}
& TH_f\exp\{-itH_f\}{\bf f} &   \\ & = \Bigl(\sum_{I_2\subset
t\mathcal O_2}\kv_{I_2}^{(f_{2,t})}\cdots\sum_{I_n\subset
t\mathcal O_n}\kv_{I_n}^{(f_{n,t})}H_\infty\sum_{I_1\subset
t\mathcal O_1}\kv_{I_1}^{(f_{1,t})}\Omega_\infty+\ldots &
\nonumber
\\ & + \sum_{I_1\subset t\mathcal
O_1}\kv_{I_1}^{(f_{1,t})}\cdots\sum_{I_{n-1}\subset t\mathcal
O_{n-1}}\kv_{I_{n-1}}^{(f_{{n-1},t})} H_\infty\sum_{I_n\subset
t\mathcal O_n}\kv_{I_n}^{(f_{n,t})}\Omega_\infty\Bigr) & \nonumber
\\ & +o(|t|^a). & \nonumber
\end{eqnarray}
By (\ref{hkk}), the first term in the r.h.s. of (\ref{ht}) equals
the expression in brackets in (\ref{th}), therefore the integrand
in (\ref{cook}) is indeed $o(|t|^a)$ for any $a$, and the wave
operators $W_\pm$ exist on $S$.

It remains to show that the wave operators are isometric. It
suffices to prove that for any ${\bf f}^{(l)}\in S_{n_l}, l=1,2,$
we have $$\lim_{t\to\infty}\langle T\exp\{-itH_f\}{\bf f}^{(1)},
T\exp\{-itH_f\}{\bf f}^{(2)}\rangle=\langle {\bf f}^{(1)}, {\bf
f}^{(2)}\rangle.$$ If $n_1=n_2=n$, then, if ${\bf
f}^{(l)}=(f_1^{(l)}\otimes\cdots\otimes f_n^{(l)})^{sym}$, the
r.h.s. equals $\sum_\sigma\prod_{k=1}^n\langle
f_k^{(1)},f_{\sigma(k)}^{(2)}\rangle$, where the sum is over
permutations of $n$ elements; otherwise the r.h.s. vanishes.

Expanding in $\xi_x$, we see that the l.h.s. equals
$$\sum_{\genfrac{}{}{0pt}{}{x_k^{(1)}\in\znu\cap t \mathcal
O^{(1)}_k}{k=1,\ldots,n_1}}
\sum_{\genfrac{}{}{0pt}{}{x_k^{(2)}\in\znu\cap t \mathcal
O^{(2)}_k}{k=1,\ldots,n_2}} \prod_{k=1}^{n_1}\langle
e^{-itH_f}f^{(1)}_k,\xi_{x^{(1)}_k}\rangle \prod_{k=1}^{n_2}
\langle \xi_{x^{(2)}_k}, e^{-itH_f}f^{(2)}_k\rangle $$
\begin{equation}\label{tlt}\times \langle T
(\xi_{x^{(1)}_1}\otimes\cdots\otimes\xi_{x_{n_1}^{(1)}})^{sym},
T(\xi_{x_1^{(2)}}\otimes\cdots\otimes\xi_{x_{n_2}^{(2)}})^{sym}
\rangle+o(|t|^a).\end{equation} Here $\mathcal O_k^{(l)}$ are open
sets for functions $f_k^{(l)}$, disjoint for equal values of $l$.
We can restrict the summation to $x_k^{(l)}\in t\mathcal
O_k^{(l)}$, with the remainder being $o(|t|^a)$ for any $a$, by
Lemma 3. For this range of $x_k^{(l)}$, using expansions for
$\xi_x$, the decay of correlations (\ref{oia}), and the fact that
the minimal distance between such points with common $l$ grows
linearly in $t$, one can establish an asymptotic orthogonality of
the vectors in the last scalar product of (\ref{tlt}): $$\langle T
(\xi_{x^{(1)}_1}\otimes\cdots\otimes\xi_{x_{n_1}^{(1)}})^{sym},
T(\xi_{x_1^{(2)}}\otimes\cdots\otimes\xi_{x_{n_2}^{(2)}})^{sym}
\rangle=\delta_{\{x^{(1)}\},\{x^{(2)}\}}+o(\gamma^{|t|})$$ with
some $\gamma<1$, where $\delta_{\{x^{(1)}\},\{x^{(2)}\}}$ equals 1
if $\{x^{(1)}_k,k=1,\ldots,n_1\}=\{x^{(2)}_k,k=1,\ldots,n_2\}$ as
sets and 0 otherwise. It follows that for $n_1\ne n_2$ the
expression (\ref{tlt}) converges to 0, and for $n_1=n_2=n$ it
converges to
$$\sum_\sigma\sum_{\genfrac{}{}{0pt}{}{x_k\in\znu}{k=1,\ldots,n}}
\prod_{k=1}^{n}\langle e^{-itH_f}f^{(1)}_k,\xi_{x_k}\rangle
\prod_{k=1}^{n} \langle \xi_{x_{\sigma(k)}},
e^{-itH_f}f^{(2)}_k\rangle = \sum_\sigma\prod_{k=1}^n\langle
f_k^{(1)},f_{\sigma(k)}^{(2)}\rangle, $$ which completes the
proof.

\section*{Acknowledgements}

It is a pleasure to thank Robert Minlos who suggested this problem
to me, and Joseph Pul\'e and Tony Dorlas for their hospitality at
UCD and DIAS. The research was supported by the Irish Research
Council for Science, Engineering and Technology.


\begin{thebibliography}{99}

\bibitem{A} Albanese, C., ``Unitary dressing transformations and
exponential decay below threshold for quantum spin systems'',
Commun. Math. Phys. {\bf 134}, 1-27, 237-272 (1990).

\bibitem{AMZ1} Angelescu, N., Minlos, R.A., Zagrebnov V.A., ``The lower
spectral branch of the generator of the stochastic dynamics for
the classical Heisenberg model''. In: Minlos, R.A., Shlosman, S.,
Suhov, Yu.M. (eds.) {\it On Dobrushin's way. From probability
theory to statistical physics.} Amer. Math. Soc. Transl. (2) {\bf
198}, (2000)

\bibitem{AMZ2} Angelescu, N., Minlos, R.A., Zagrebnov, V.A., ``The
one-particle energy spectrum of weakly coupled quantum rotators'',
J. Math. Phys. {\bf 41}, 1-23 (2000)

\bibitem{BR}Bratteli, O., Robinson, D.W.: {\it Operator Algebras and Quantum
Statistical Mechanics}, 2nd ed.,  Springer Verlag, Berlin, vol. 1,
1987, vol. 2, 1996

\bibitem{DK} Datta, N., Kennedy, T., ``Expansions for one quasiparticle states
in spin 1/2 systems'', J. Stat. Phys. {\bf 108}, 373-399 (2002)

\bibitem{H1} Haag, R., ``Quantum field theories with composite
paticles and asymptotic completeness'', Phys. Rev. {\bf 112},
669-673 (1958)

\bibitem{H2} Haag, R., ``The framework of quantum field theory'', Nuovo
Cimento Supp. {\bf 14}, 131-152 (1959)

\bibitem{I} Iarotski, D.A.,  `` `Free' evolution of multi-particle
excitations in the Glauber dynamics at high temperature'', J.
Stat. Phys., {\bf 104}, 1091-1111 (2001)

\bibitem{KT1} Kennedy, T., Tasaki, H., ``Hidden $Z_2\times Z_2$ symmetry breaking in Haldane gap
antiferromagnets'', Phys. Rev. B {\bf 45}, 304 (1992)

\bibitem{KT2} Kennedy, T., Tasaki, H., ``Hidden symmetry breaking and the Haldane phase in $S=1$ quantum
spin chains'', Commun. Math. Phys. {\bf 147}, 431-484 (1992)

\bibitem{KT} Kirkwood, J. R., Thomas, L. E., ``Expansions and phase
transitions for the ground states of quantum Ising lattice
systems'', Commun. Math. Phys. {\bf 88}, 569-580 (1983)

\bibitem{KM} Kondratiev, Yu.G., Minlos, R.A., ``One-particle subspaces in the stochastic
XY model'', J. Stat. Phys. {\bf 87}, 613-642 (1997)

\bibitem{Ma1} Malyshev, V.A., ``One-particle states and scattering
theory for Markov processes'', in: {\it Locally interacting
systems and their applications in biology}, Pushchino, 1976, pp.
173-193, Lecture Notes in Math., 653, Springer, Berlin, 1978

\bibitem{Ma2} Malyshev V.A., {\it Elementary introduction to the
mathematical physics of infinite-particle systems} [in Russian],
Lectures for Young Scientists, R17-83-363, JINR, Dubna, 1983

\bibitem{MM2} Malyshev V. A.,  Minlos R. A.: {\it Linear operators in infinite
particle systems}, AMS, Providence, RI, 1995

\bibitem{M1} Matsui, T., ``A link between quantum and classical Potts
models'', J. Stat. Phys. {\bf 59}, 781-798 (1990)

\bibitem{M2} Matsui, T., ``Uniqueness of the translationally invariant
ground state in quantum spin systems'', Commun. Math. Phys. {\bf
126}, 453-467 (1990)

\bibitem{M} Minlos, R.A., ``Invariant subspaces of the stochastic Ising high temperature
dynamics'', Markov Processes Relat. Fields {\bf 2}, 263-284 (1996)

\bibitem{MS} Minlos, R.A., Suhov, Yu.M., ``On the spectrum of the generator
of an infinite system of interacting diffusions'', Commun. Math.
Phys. {\bf 206}, 463-489 (1999)

\bibitem{MT} Minlos, R.A., Trishch, A.G., ``Complete
spectral resolution of the generator of Glauber dynamics for the
one-dimensional Ising model'',  Russian Math. Surveys {\bf 49},
no. 6, 210-211 (1994)

\bibitem{RS3} Reed, M., Simon, B.: {\it Methods of modern mathematical
physics.} v.3: {\it Scattering theory.} N.Y.: Academic press, 1979

\bibitem{R} Ruelle, D., ``On the asymptotic condition in quantum
field theory'', Helv. Phys. Acta {\bf 35}, 147-163 (1962)

\bibitem{Y1} Yarotsky, D.A., ``Perturbations of ground states in
weakly interacting quantum spin systems'', J. Math. Phys. {\bf
45}, 2134-2152 (2004)

\bibitem{Y2} Yarotsky, D.A., ``Uniqueness of the ground state in weak
perturbations of non-interacting gapped quantum lattice systems'',
to appear in J. Stat. Phys. {\bf 118}, 119-144 (2005) 

\bibitem{Y3} Yarotsky, D.A., ``Scattering of quasi-particle excitations in weakly coupled
stochastic lattice spin systems'', Comm. Math. Phys. {\bf 249},
449-474 (2004)

\bibitem{ZhKM} Zhizhina, E.A., Kondratiev, Yu.G., Minlos, R.A., ``The lower branches of the
Hamiltonian spectrum for infinite quantum systems with compact
`spin' space'', Trans. Moscow Math. Soc. {\bf 60}, 225 (1999)



\end{thebibliography}
\end{document}